# The effect of surfactants and their concentrations on the liquid-exfoliation of graphene

Shuai Wang[1], Min Yi*[1], Shuaishuai Liang[1], Zhigang Shen*[1,2], Xiaojing Zhang[1] and Shulin Ma[1]

[1] Beijing Key Laboratory for Powder Technology Research and Development, Beijing University of Aeronautics and Astronautics, Beijing 100191, China

[2] School of Materials Science and Engineering, Beijing University of Aeronautics and Astronautics, Beijing 100191, China

*Corresponding Author

Email: yi@mfm.tudarmstadt.de; shenzhg@buaa.edu.cn

Address: Beijing University of Aeronautics and Astronautics, 37 Xueyuan Road, Beijing 100191, China

Phone: 86-10-82317516

Fax: 86-10-82338794

**Abstract**

We investigated the effect of surfactants and their concentration ($C_{sur}$) on the final graphene concentration ($C_G$) via the liquid-phase exfoliation method. Six typical surfactants including ionic and non-ionic ones were explored and the optimized $C_{sur}$ for each surfactant was suggested. For ionic surfactants, $C_G$ increases with $C_{sur}$ before reaching its maximum and then maintains the high level. The different mechanisms of ionic and non-ionic surfactants in stabilizing graphene dispersions are explained by the theory for colloidal stability. The as-prepared graphene sheets are verified to be highly exfoliated through transmission electron microscopy and atomic force microscopy studies, while the defect-free structure was evidenced by Raman spectra and X-ray photoelectron spectroscopy.

# 1. Introduction

Graphene with its unique 2-dimentional (2D) structure has gained great interests since it was first discovered in 2004.[1,2] It has large specific surface area (theoretical 2630 $m^2g^{-1}$, experimental 400-700 $m^2g^{-1}$)[3], high Young's modulus (~1.0 TPa), high optical transmittance (~97.7 %)[4], high intrinsic mobility (200 000 $cm^2v^{-1}s^{-1}$)[5] and with the thermal conductivity of about 5000 $Wm^{-1}K^{-1}$.[6] These excellent mechanical[7], optical[8], electrical[9] and thermal[10] properties lead to a research hotspot of synthesizing graphene of high quality (larger size and fewer layers) in more efficient, green and industrial way.[11]

Mechanical exfoliation and epitaxial growth can produce graphene with high quality, but high cost and low throughput limit their application areas.[12] Chemical vapor deposition (CVD) method of synthesizing few layer graphene (FLG) was first discovered in 2006[13], and to date, mature millimeter-sized monolayer graphene sheets can be achieved and transferred to many substrates by plasma-enhanced CVD process.[14] But direct growth of large-sized graphene and on different substrates still need to be explored and industrial-scale synthesis is to be solved. Chemical reduction of graphene oxide (rGO) was thought to be a promising approach, as graphene oxide (GO) can be easily exfoliated through sonication in liquid phase,[15,16] and chemical reduction process can decrease the functional defects produced by oxidation.[17] But many structural defects cannot be removed from the reduction process and thus degrade the excellent electronic properties that set graphene apart from traditional materials.[18-22]

Apart from above methods, direct exfoliation of graphite in liquid phase opens a door of industrial synthesizing graphene with simplicity and efficiency since Coleman et al. first accomplished it in 2008.[23] Exfoliation in organic solvent[24], with polymer assistant[25] and with surfactant assistant[26] are three different routes which can enhance the exfoliation process. In spite of some defects brought by surfactant, the last one was proved to be an ideal way of preparing high-concentration and strong-stability graphene dispersion. Ionic surfactants were first introduced to assist the exfoliation process. Valiyaveettil et al. used a cation surfactant cetyltrimethylammonium bromide (CTAB)[27], Coleman et al. used sodium dodecyl benzene sulfonate (SDBS)[28], sodium cholate (SDOC)[26] and other ten kinds of surfactants[29] to help exfoliating graphite and promising $C_G$ was achieved. Guardia et al. explored the differences between ionic and non-ionic surfactants in assisting exfoliation and verified the ascendency of non-ionic surfactant.[30] Du et al. introduced some organic salt to assist exfoliation and enhanced the efficiency up to 123 times.[31]

In order to direct the innovation in enhancing the liquid phase exfoliation effect, two significant issues must be addressed: on one hand, how many factors influence the exfoliation degree (input); on the other hand, which index can represent the exfoliation degree (output). Much attention was paid to these questions by researches worldwide. As for the first question, based on the predecessor's works, some particular factors, i.e., sonication time ($t_{sonic}$), centrifugation (CF) speed and initial graphite concentration ($C_{Gi}$) were discussed as a function of $C_G$. For ionic (of zeta potential) and non-ionic (of steric effect) surfactant, there are different mechanisms of stabilizing graphene dispersion[29]; 5 hours sonication may have a decent marginal benefit[30]; The increase of CF will negatively affect $C_G$ and graphene sheet's quantity[26]; $C_G$ equals a factor (0.01 for SDBS) multiplied by the square root of $C_{Gi}$[32]. But to the best of our knowledge, the effectiveness of $C_{sur}$ on $C_G$ was not deeply discovered. As for the second question, Coleman et al. first use absorption of the dispersion as a main index for exfoliation according to the Lambert-Beer Law and use transmission electron microscope (TEM) and other characterization tools to examine the quality of the dispersion. Whether graphene dispersion with high $C_G$ has an analogous quality to the one with relatively low concentration is not solidly confirmed.

In this paper, six kinds of surfactants were used to probe the $C_{sur}$ as a function of $C_G$ and preliminarily discuss the mechanism of the result. The best concentration of each surfactant for exfoliation was found. We introduce two models to explain the differences between ionic and non-ionic surfactants in the sedimentation process. In order to get a fully understanding of the factors that influence the final $C_G$, many controlled experiments were employed. Some characterization methods were performed to examine the quality (sheet size, number of the layers and structural defects) of the product and some interesting phenomena were found. The results provide a valuable reference for graphene exfoliation in water/surfactant dispersion.

## 2. Experimental

**1.1 Materials**

The natural graphite flakes was purchased from Alfa Aesar Co., Ltd. (-325mesh, 99.8%). SDOC and other surfactants were purchased from Sinopharm Chemical Reagent Co., Ltd. The purified water was purchased from Beijing Kebai'ao Biotech. Co., Ltd. All the materials employed in the experiments were used as received.

**1.2 Exfoliation process**

Here we take SDOC for example, and other surfactants are processed with the same method. First, graphite dispersion was prepared by adding 1 g graphite powder to 200ml SDOC/water mixture in 300 ml capped round-bottomed flask ($C_{Gi}$=5 mg/mL). The solvent was prepared by adding different quantities of SDOC in purified water, by which we can tune the SDOC concentration, i.e. 0.025 mg/ml, 0.05 mg/ml, 0.1 mg/ml, 0.25mg/ml, 0.5 mg/ml, 1 mg/ml and 2.5 mg/ml. It is worth noting that the

SDOC in water becomes hard to dissolution as the concentration of SDOC rise, so ultrasonication for about 1 minute was employed to accelerate the dissolution. The SDOC solution mixed with graphite was transferred into six same-sized reagent bottles (30 ml). All the graphite solutions were then ultrasonicated in 100 W ultrasonic bath (KX-1730T Shenzhen Kexi Chemical Co., Ltd) for 8 hours. In order to wipe off the massive graphite sediment, the bottles would stay overnight. Subsequently, the supernatant liquor was carefully transferred 10 ml test tube for CF. Then the centrifuged liquor was extracted for the measurement of absorption by a UV-visible light spectrophotometer (TV-1900 Beijing Purkinje General Instrument Co., Ltd.; 660 nm wave length), through which we can determine the concentration of graphene dispersed in the solution. According to the Lambert-Beer Law, $A=\alpha_{660nm}C_G l$, $C_G$ can be obtained from absorption with the coefficient of $\alpha_{660nm}=1458$ $mLmg^{-1}m^{-1}$ (see results and discussion section). Different from powder-like SDOC, Tween 80 is a kind of sticky liquid, so that magnetic stirring for 10h was need before graphite was added to the solution.

### 1.3 Characterization method

TEM and high-resolution TEM (HRTEM) images were performed by a JEOL JEM-2010FEF operated at 200 kV. Atomic force microscope (AFM) images were purchased by a Multimode 8 microscope (Bruker Corporation) equipped with a ScanAsyst-Air probe in ScanAsyst mode. Raman spectroscopy was performed on a Renishaw inVia Raman microscope with a 532 nm He–Ne laser. X-ray diffraction (XRD) spectra were collected by Rigaku D/max 2200 X-ray diffractometer at 40kV and 40mA. X-ray photoelectron spectroscopy (XPS) spectra were obtained by ThermoFisher Scientific's ESCALAB, with a 150 W monochromated Al Kα X-ray source.

TEM samples were prepared by pipetting a few micro liters of the dispersion onto holey carbon mesh grid. For AFM sample preparation, we first diluted the centrifuged supernatant clear solution by acetone (1:1000), and then ultrasonicated the solution for about 1minute. One drop of solution was dripped onto the mica substrate. The prepared sample was dried in ambient temperature. Since Tween 80 and Triton X-100 have poor compatibility with organic solvents, the solution with Tween 80 was diluted by deionized water. Raman samples were obtained by filtering graphene dispersion on an organic membrane filter, followed by drying in ambient conditions. The Raman, XRD and XPS samples were prepared by filtering graphene dispersion onto the organic filter membrane.

## 3. Results and discussion

**1. Five factors influence $C_G$**

$C_G$ was determined through absorption, and calculated by Beer-Lambert law.[23] Water/surfactant/graphene dispersion spectrum in the range from 300 nm to 900 nm shows that 660 nm is a decent wavelength for characterization. According to Lambert-Beer Law, Absorption of the dispersion equals to a factor $\alpha_{660nm}$ multiplies $C_G l$. As $\alpha_{660nm}$ was valued differently by different groups[26, 28, 32], a uniform value of $\alpha_{660nm}$ must be determined. We first prepare graphene dispersion through liquid-phase exfoliation, and measure the absorption of the dispersion with the wave length of 660 nm. The dispersion (with the volume of V) was then suction filtrate with a pre-weighted membrane ($m_1$). The membrane was then vacuum dried and weighted ($m_2$). $C_G$ is determined by the formula *$C_G=(m_1-m_2)/V$*.

Fig. 1 shows A/l as a function of $C_G$, and $\alpha_{660nm}$ can be confirmed by the slope of the fitting straight line. The dispersion with high $C_{sur}$ ($C_{sur}$=5 mg/mL) has a relatively low value of $\alpha_{660nm}$ (578.5 mg/mL/m), while dispersion with low $C_{sur}$ ($C_{sur}$=0.1 mg/mL) has a relatively high value of $\alpha_{660nm}$ (1458.3 mg/mL/m). This result should be attributed to the fact that for dispersion with high $C_{sur}$, surfactant cannot be totally washed off during the filtration process. $(m_2-m_1)$ is the sum of the mass of surfactant and graphene, so the value of C is larger, and the value of $\alpha_{660nm}$ is smaller than the actual value. Hence, $\alpha_{660nm}$ of low $C_{sur}$ dispersion, i.e. 1458.3 mg/mL/m, is recommended as the reference value for various kinds of surfactant concentration.

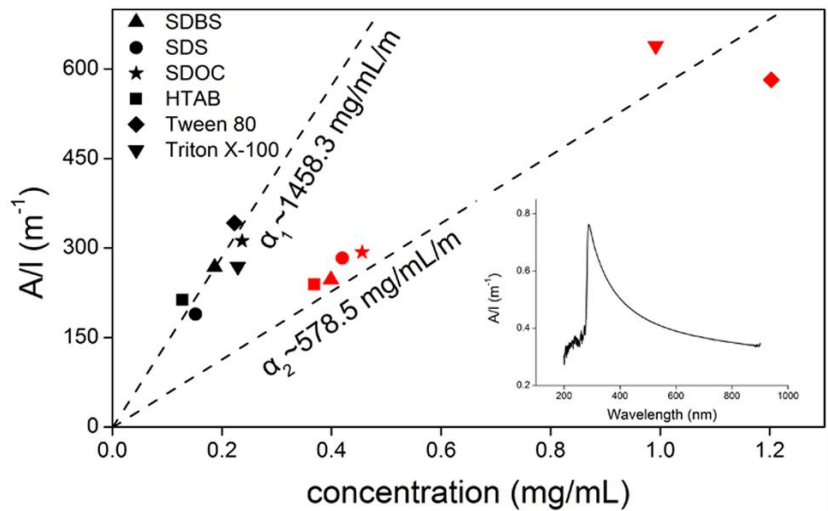

**Fig. 1** Absorption per unit length as a function of graphene dispersion concentration. Each dot of different shape represents a kind of surfactant. Black dots represent that $C_{sur}$ is 0.1mg/mL and red dots represent that $C_{sur}$ is 5 mg/mL. Inset: UV-visible spectrum of typical graphene dispersion from 200 nm to 900 nm ($C_{Gi}$=5 mg/mL, $t_{sonic}$= 8h, CF: 1500 rpm for 30 min).

Five vital factors, i.e., $t_{sonic}$, $C_{Gi}$, CF rate, surfactant type and $C_{sur}$ mainly influence $C_G$. In order to get a full view of $C_G$'s dependence on $C_{sur}$, other factors must be pre-discussed. We first use dispersion with different $C_{Gi}$ to do the experiment. From the 3D bars in Fig. 2a and Fig. 2b we can find that, both $C_{Gi}$ and $C_{sur}$ affect $C_G$: for low $C_{Gi}$, i.e. 1 mg/mL, $C_G$ reaches its peak at a low $C_{sur}$ (SDOC: 0.1 mg/mL, Tween 80: 0.5 mg/mL); for high $C_{Gi}$, i.e., 10 mg/mL, $C_G$ reaches its peak at a relatively high $C_{sur}$

(SDOC: 1 mg/mL, Tween 80: 1 mg/mL). It is understandable, since larger numbers of graphite flakes in the dispersion needs more surfactant molecules to adhere on. So the best $C_{sur}$ for exfoliation is related to $C_{Gi}$. Hence, it is not appropriate for the previous works to emphasize the best $C_{sur}$ for exfoliation, neglecting the role of $C_{Gi}$.

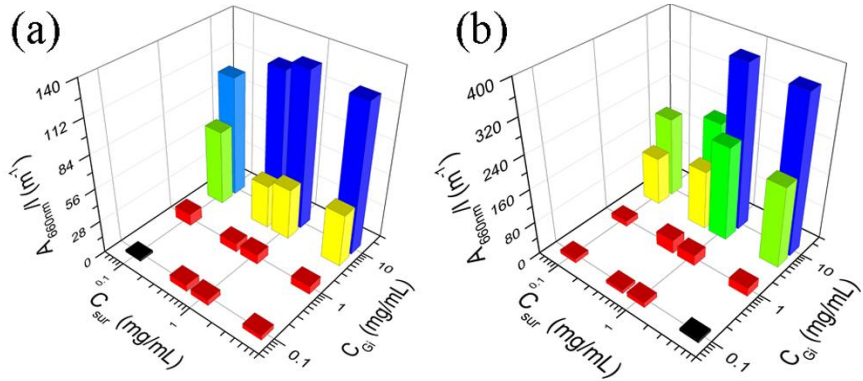

**Fig.2** 3D bar of graphene dispersion concentration as a function of $C_{sur}$ and $C_{Gi}$ with the assistance of (a) SDOC (b) Tween 80 ($t_{sonic}$=8 h, CF: 1500 rpm for 30 min).

We then consider $t_{sonic}$, which may likely affect $C_G$. The $C_G$-T curve of the six surfactants was depicted in Fig. 3a. $C_{Gi}$ is uniformly 0.1 mg/mL. From the curves, we can find that absorption approximately linearly increased as time passes. The input energy increases with $t_{sonic}$, and so does the exfoliation degree. It is worth noting that after 8 hours sonication, the increase of absorption slows down. After long time sonication (8 h), most of the surfactants were adhered to the graphene sheets. The decrease of $C_{sur}$ may occur after this process, and the increase of $C_G$ may retard. In order to minimize the influence of $t_{sonic}$ on $C_G$, for the following experiments, $t_{sonic}$ was all fixed at 8 h.

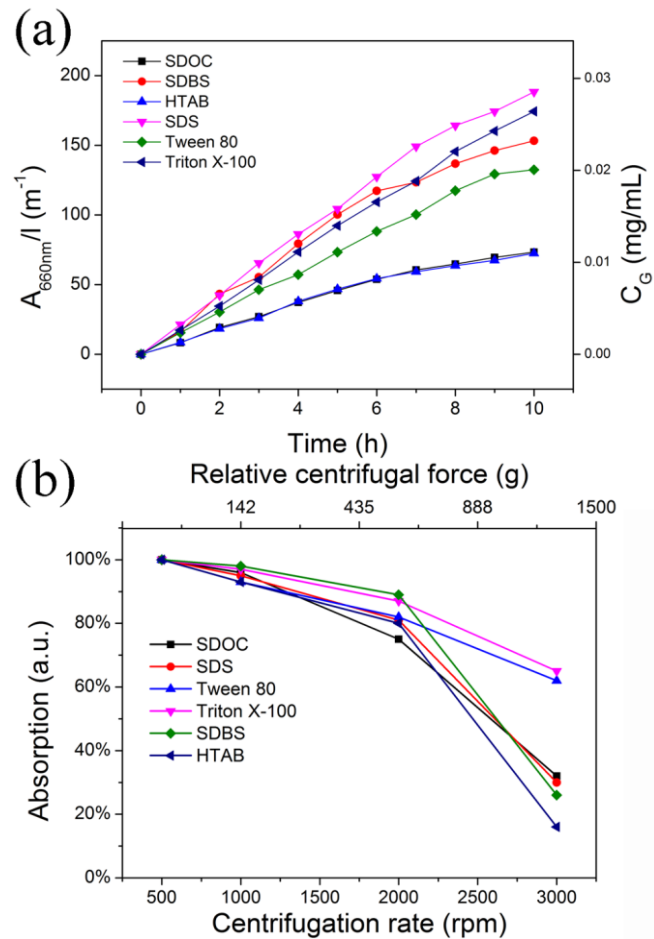

**Fig.3** (a) Absorption per unit length of graphene dispersion, $A_{660nm}/l$, as a function of time with six kinds of surfactants. (b) Relative absorption as a function of centrifugal force, which was conversed from CF rate.

CF rate also determines the graphene dispersion in both concentration and quality aspects. The same samples (SDBS, $t_{sonic}$=8 h, $C_{Gi}$=5 mg/mL) were processed with different CF rate and time, and the results were shown in Fig. 3b. Higher CF rate leads to higher quality graphene dispersions, and vice versa. Long time CF can also decrease $C_G$. For CF higher than 2000 rpm, much of the graphene sheets and graphite flakes will precipitate. Based on above discussions, 1500 rpm CF is suggested. To sum up, 8 h sonication, 1500 rpm CF may have a decent effect on producing graphene dispersion; meanwhile the interaction between $C_{sur}$ and $C_{Gi}$ should be considered.

## 2. $C_G$ as a function of $C_{sur}$

Based on above conclusions, we designed an experiment to probe the impact of surfactant with different concentration to help exfoliating graphite powder in water/surfactant solution, as shown in Fig 4. The critical micelle concentrations (CMC) of each surfactant were depicted in red dash line. CMC was originally thought as the minimum $C_{sur}$ required for successful dispersion of graphite, but denied by other experiments.[26] The comparison between the optimum $C_{sur}$ and the CMC of the surfactant in Fig. 4 find no clues of their connections. Each ionic surfactant has a best $C_{sur}$ for exfoliation, i.e., 0.1 mg/mL for SDOC, 0.05 mg/mL for SDBS, 0.5 mg/mL for SDS and 0.5 mg/mL for HTAB.

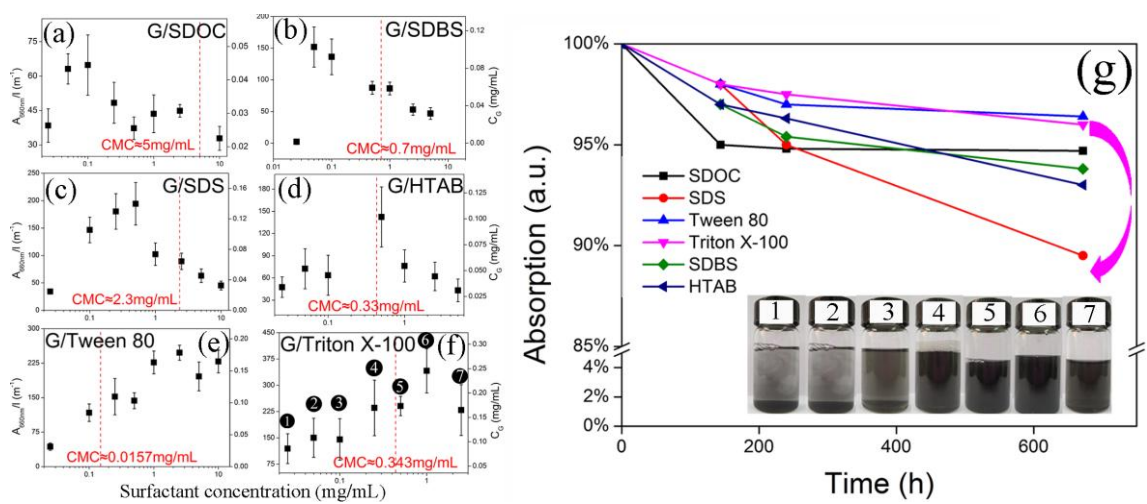

**Fig. 4** (a)-(f) Optical absorbance per unit length ($A_{600}$nm/l) and $C_G$ as a function of $C_{sur}$ ($t_{sonic}$=8 h, CF: 1500 rpm for 30 min). CMC of each surfactant was shown by the vertical red dash line. (g) Sedimentation curves for six surfactants after CF. inset: graphene dispersion after CF using Triton X-100 as the surfactant. $C_{sur}$ from left to right: 0.025 mg/mL, 0.05 mg/mL, 0.1mg/mL, 0.25 mg/mL, 0.5 mg/mL, 1 mg/mL and 5 mg/mL.

An interesting phenomenon was found that, as $C_{sur}$ growing, different types of surfactants performed differently and $C_G$ curves show different tendencies. For ionic surfactants, i.e., SDOC, HTAB, SDS and SDBS, $C_G$ reaches its climax, and then decreases. For non-ionic surfactants, i.e., Tween 80 and Triton X-100, $C_G$ reaches its peak value and remains with the increase of $C_{sur}$. Such results may be explained by the fact that ionic surfactant and non-ionic surfactant have different mechanisms for stabilizing the colloid according to the theory of Coleman's group. For ionic surfactant, graphite flakes were first exfoliated by sonication-derived cavitation and shear force, and the exfoliated graphene sheets were then adhered by charged surfactant molecules. When the two charged graphene sheets approaches each other, Derjaguin-Landau-Verwey-Overbeek (DLVO) theory can be used to explain the anti-aggregation mechanism.[33] The potential energy per unit area between two infinitely extended solids with the gap of x can be calculated by:

$$w(x) = 64 c_0 k_B T \lambda_D \cdot \tanh^2\left(\frac{e \psi_0}{4 k_B T}\right) \cdot e^{-x/\lambda_d} - \frac{A_H}{12 \pi x^2} \quad [34]$$

in which, $\Psi_0$ represents the surface potential, $C_0$ represents the bulk concentration of the salt, $k_B$ represents the Boltzmann constant, T represents the local temperature, $\lambda_D = \kappa^{-1}$ represents the decay length, $\kappa = \sqrt{(2C_0 e^2 / \varepsilon \varepsilon_0 k_B T)}$. $A_H$ represents the Hamaker constant $A_H = \pi^2 C_{AB} \rho_A \rho_B$.

The first part of the formula stands for the double layer electrostatic repulsion force, while the last part of the formula stands for the van der waals attraction force (vdw force). As the salt concentration rises, the potential barrier $V_{BD}$ decrease ($A_H$ increased much with the salt concentration, so does the vdw force), and the graphene sheets are easier to aggregate and the dispersion become less stabilized. Therefore, we can explain the trend of the ionic surfactant-assisted exfoliation in Fig. 4. At low $C_{sur}$ condition, $C_G$ increases with $C_{sur}$, because the exfoliated graphene sheets are in great demand for surfactant molecules to adhere. The addition of surfactant can enhance the stabilization of the dispersion. At relatively high $C_{sur}$, the excess of the surfactant molecules will decrease $V_{BD}$, and $C_G$ will be lower after CF.

For non-ionic surfactants, the graphene sheets adhered by the surfactant molecules were merely charged, and the double layer electrostatic force becomes much lower. When the hydrophobic tails (hydrocarbon chain) of the surfactant molecules from two coated graphene sheets interact, the steric repulsion force plays a more important role. Steric force was used to explain the lack of adhesion or aggregation of uncharged lipids, and it was proved to be dominating over the electrostatic force and vdw force in a short distance.[35] De Gennes established a model to calculate the interaction force and corresponding energy between two polymer-coated sheets, and the force per unit area was given by,

$$F = \frac{k_B T}{D^3} \left[ \left(\frac{h_c}{h}\right)^{9/4} - \left(\frac{h_c}{h}\right)^{3/4} \right] \quad 36$$

in which, $h_c$ represent the chain length of the adhered molecule which adhered to the sheet, D represent the average distance between two junction points which connect the sheet and the adhered molecule. The first term in the bracket represents osmotic force; the second one represents the elastic force. At strong compressions, the osmotic term should dominate completely. We can find that the interaction force is in inverse proportion to the third power of D. As $C_{sur}$ increases, more surfactant molecules were attached onto the exfoliated graphene sheets, and filled the surface, which lead to a decrease of D. Stronger repulsion force F will avoid graphene sheets from aggregation. When the surface of the graphene sheets were filled up, the effect of surfactant molecules ceased, but the sheets still being separated, so $C_G$ will not fall down. Sedimentation curves show great stability of the as-made graphene dispersion. Over

80% of the graphene remained in the dispersion after 20-day standing.

## 3. Characterization

The exfoliated graphene sheets were characterized by TEM, AFM, and Raman spectra to examine their quality. Fig. 5 shows typical TEM images of the exfoliated graphene flakes. The edge of the graphene in Fig. 5a is a protruding few-layer graphene. From the edge we can count the number of the flake layer. Fig. 5b shows the wrapped edge of 5a, from which the few-layer graphene sheets can be easily found. The HRTEM picture in 5c shows that the graphene sheet in 5b has the sheet number less than five. The stripe in Fig. 5d Graphene shows the existence of graphene nano ribbon (GNR).

The sheets have also been verified by AFM, which is an easy way to detect the thickness of the graphene sheets. Fig. 6a shows some graphene sheets with the thickness of ~1 nm (surfactant: Triton X-100). As shown in Fig. 6b, large numbers of graphene sheets could be seen, with the average area of 46.83 $\mu m^2$ and thickness of ~1 to 3 nm (surfactant: SDOC). In Fig. 6a and 6b, the white dots on the graphene flakes and the mica substrate could be the agglomerated surfactant molecules after water was vaporized away. As shown in Fig.6, shapes of the white dots and the graphene sheets in two figures are different. The shape of the sheets is cotton-shaped in Fig. 6a, while the shape is uniformly block-shaped in Fig. 6b. The reason for this phenomenon is still not clear yet.

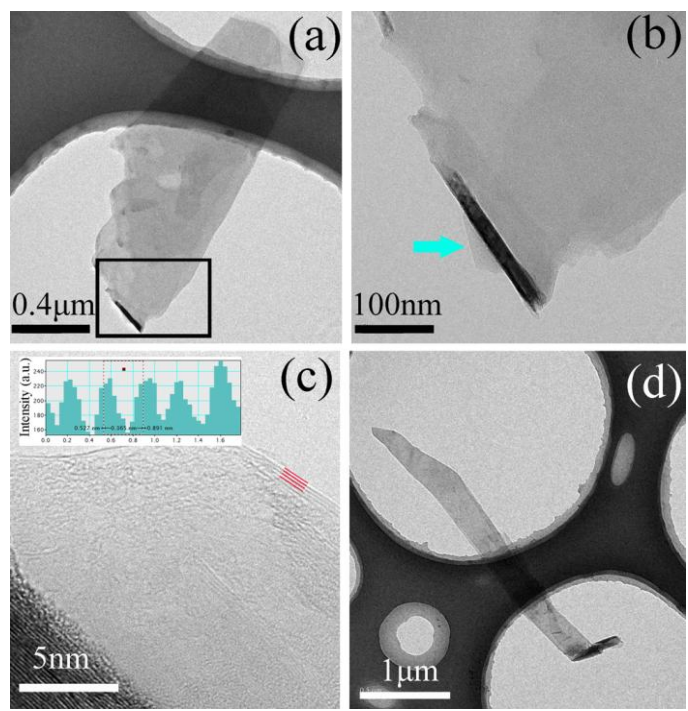

**Fig. 5** (a), (b) TEM images of one graphene flake. The cyan arrow indicates a graphene sheet protruding from a curly sheet. (c) A HRTEM image of the edge area in (b) Inset: Intensity profile recorded in the region of the marked red line showing the edge fringe separation of~0.37 nm. (d) An image of a GNR.

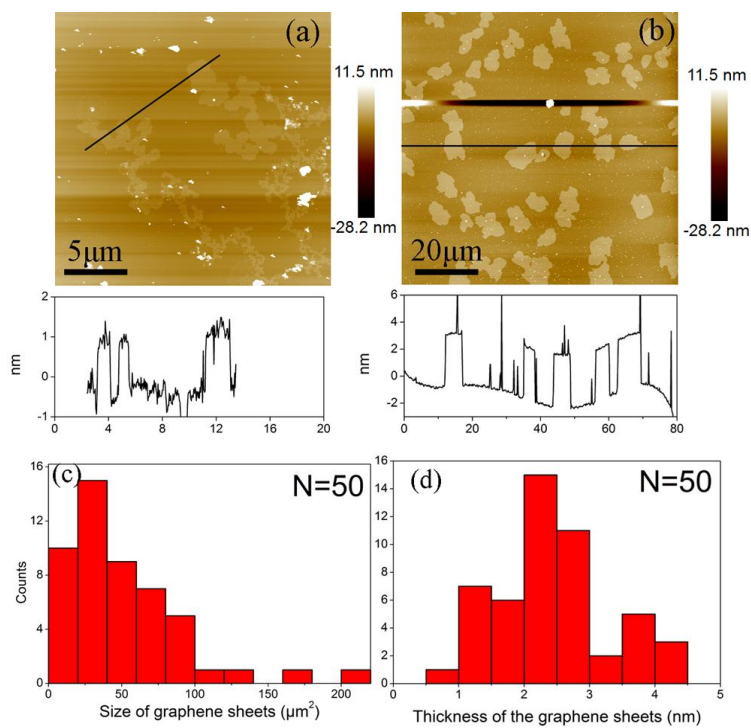

**Fig. 6** (a), (b) AFM images of graphene sheets with the assistance of (a) Triton X-100, (b) SDOC. (c), (d) Histogram of size and thickness of the graphene sheets, respectively.

Besides TEM and AFM images, Raman spectra of the graphite powder and the filtered film prepared from graphene dispersions are presented in Fig. 7a, from which we can measure the defect degree of the as-produced graphene. The graphene flakes may mainly suffered edge defects rather than basal-plane disorder defects[26], in view that in the filtered film the D band is relatively weak and the G band is not broadened, whereas disorder defects in basal plane often result in much higher D band and largely broadened G band as commonly found in GO or chemical reduced graphene. The shape of the 2D band of the filtered film is intrinsically different from that of graphite, indicating exfoliation state of the as-produced graphene.[37] Meanwhile, its relatively high intensity with respect to G band ($I_{2D}/I_G$), also gives a strong evidence of the existence of few-layer graphene flakes.

The XRD spectra supports the AFM results that graphene sheets with less than 5 layers were obtained from sonication in liquid phase as shown in Fig. 7b. Compared with pristine graphite filtered film, the exfoliated graphene sheets shows very weak peak appeared at 2θ-26.6 ° corresponding to the (002) planes, which is symbolic for graphite powder. Hence, we can draw a conclusion that, after exfoliation, the distance between $sp^2$ hybrid constructed carbon layers was not changed, but the number of this layer to layer gap was decreased. Furthermore, no (004) peak can be found, which indicate that long-range order greater than four layers were eliminate by exfoliation procedure.

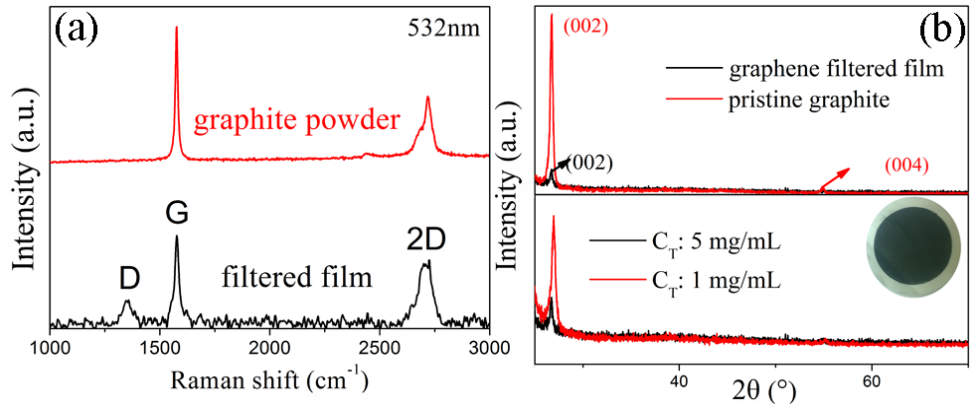

**Fig. 7** (a) Comparison of Raman spectra between graphite powder and filtered graphene films. (b) Upper: XRD spectra of graphene filtered film (SDOC 5 mg/mL, $t_{sonic}$ =8 h, CF: 1500 rpm, 30 min) and pristine graphite filtered film. Lower: The comparison of XRD spectra of graphene filtered film with different surfactant (Triton X-100) concentration. *Inset* a photograph of a filtered graphene film.

In Fig. 4, we can find that graphite/water dispersion with 5 mg/mL Triton X-100 addition has a relatively poor performance on the final concentration compared with 1 mg/mL Triton X-100 addition. $C_G$, an index commonly used for evaluation and to instruct the exfoliation method. Fig. 7b shows the XRD spectra of graphene filtered films with different Triton X-100 concentrations, Where the (002) peak of 1mg/mL is relatively higher than the 5 mg/mL case. The gap number between layers is smaller for 1 mg/mL than 5 mg/mL dispersion. And (004) peak can be found in red line, even though it is very small. All these clues indicate that although $C_G$ is higher for 1 mg/mL Triton X-100 addition than 5 mg/mL, the latter one is more effective in the exfoliation process.

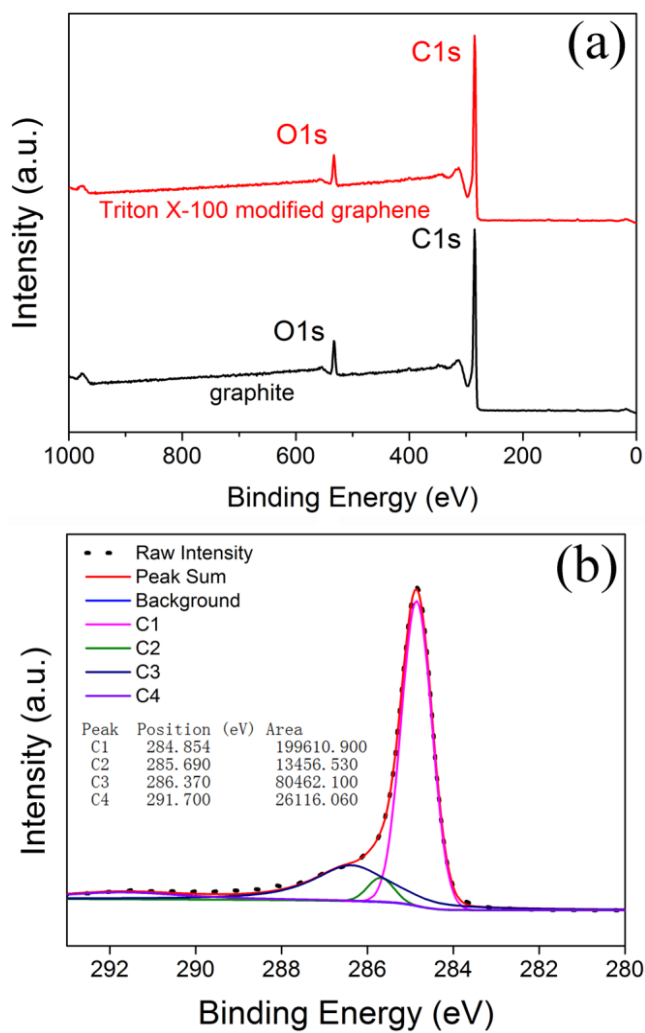

**Fig.8** (a) XPS survey of graphite flakes and graphene from surfactant assisted exfoliation. (b) XPS narrow scan of graphene in C1s range.

XPS survey of graphene sheets shows dominating C atom content compared to O atom (97.09% of C atoms to 2.91% of O atoms; Triton 5 mg/mL), which indicate that surfactants with high concentration didn't oxidize the graphite flakes during the exfoliation process. In Fig. 8a, XPS spectrum of graphene presents similarity to that of graphite. Considering the structure of Triton X-100, we fit the C1s peak with binding energies at ~284.80 eV ($sp^2$ C, C1), ~285.69 eV ($sp^3$ C, C2), ~286.37 eV (C in C-O-C, C3), and 291.7 eV (C in π-π bond, C4) as shown in Fig. 8a. C1 is related to the graphene structure, while C2, C3 and C4 are related to Triton X-100's structure.

The areas under C2 and C4 curve have the ratio of 1:2, which is consistent with the structure of Triton X-100. C3, which represents the C atoms consisted in the C-O-R chain in Triton X-100 also suits well. The non-$sp^2$ C atoms are in consistence with Triton X-100 structure, so C atoms in graphene are mainly $sp^2$ hybridized. Non chemical modified clues were found through XPS characterization.

## 4. Conclusions

In conclusion, we have demonstrated the influence of $C_{sur}$ on final graphene dispersion concentration. The best concentrations of six surfactants for exfoliation were measured and show little relation to the CMC of the surfactant. For ionic surfactants, $C_G$ reaches its maximum with the increase of $C_{sur}$ and then falls down; for non-ionic surfactants, $C_G$ reaches its maximum and maintains the high level. The different trends of $C_{sur}$-$C_G$ curves between ionic and non-ionic surfactant inspired us to build models to simulate the sedimentation process, and the phenomenon was successfully explained. The as-prepared graphene sheets were proved to be large sized and few-layered. These results will give us guidance in the future graphene preparation in water/surfactant medium.

**Acknowledgement:** This work was supported by Beijing Natural Science Foundation (Grant No. 2132025), the Special Funds for Co-construction Project of Beijing Municipal Commission of Education, the Fundamental Research Funds for the Central Universities in China, the Specialized Research Fund for the Doctoral Program of Higher Education (20131102110016), and the Innovation Foundation of BUAA for Ph.D. Graduates (YWF-14-YJSY-052).